\documentclass[a4paper,12pt]{article}

\usepackage[english]{babel}
\usepackage[utf8x]{inputenc}
\usepackage[T1]{fontenc}

\usepackage[a4paper,top=3cm,bottom=2cm,left=3cm,right=3cm,marginparwidth=1.75cm]{geometry}

\usepackage{setspace}
\usepackage{amssymb}

\usepackage{cite}

\usepackage{amsmath}
\usepackage{graphicx}
\usepackage[colorinlistoftodos]{todonotes}
\usepackage[colorlinks=true, allcolors=blue]{hyperref}

\title{Mirror-backed dielectric metasurface sensor with ultrahigh figure of merit based on super-narrow Rayleigh anomaly}

\author{Yonghang Wang$^{1,2}$, Lei Xiong$^{2,3}$, Ming Tian$^{1,4}$, and Guangyuan Li$^{2,5,*}$}

\date{}
\begin{document}
\maketitle

\begin{spacing}{2.0}

\noindent \large$^1$Wuhan Research Institute of Posts and Telecommunications, Wuhan 430070, China

\noindent $^2$CAS Key Laboratory of Human-Machine Intelligence-Synergy Systems, Shenzhen Institute of Advanced Technology, Chinese Academy of Sciences, Shenzhen 518055, China

\noindent $^3$School of Information Science and Engineering, Yunnan University, Kunming 650500, China

\noindent $^4$Wuhan Ligong Guangke Co. Ltd., Wuhan 430223, China

\noindent $^5$Shenzhen College of Advanced Technology, University of Chinese Academy of Sciences, Shenzhen 518055, China


\noindent *Corresponding author: gy.li@siat.ac.cn

\end{spacing}

\newpage

\begin{abstract}
Plasmonic nanostructures with large local field enhancement have been extensively investigated for sensing applications. However, the quality factor and thus the sensing figure of merit are limited due to relatively high ohmic loss. Here we propose a novel plasmonic sensor with ultrahigh figure of merit based on super-narrow Rayleigh anomaly (RA) in a mirror-backed dielectric metasurface. Simulation results show that the RA in such a metasurface can have a super-high quality factor of 16000 in the visible regime, which is an order of magnitude larger than the highest value of reported plasmonic nanostructures. We attribute this striking performance to the enhanced electric fields far away from the metal film. The super-high quality factor and the greatly enhanced field confined to the superstrate region make the mirror-backed dielectric metasurface an ideal platform for sensing. We show that the figure of merit of this RA-based metasurface sensor can be as high as 15930/RIU. Additionally, we reveal that RA-based plasmonic sensors share some typical characteristics, providing guidance for the structure design. We expect this work advance the development of high-performance plasmonic metasurface sensors.
\end{abstract}

\newpage

\section{Introduction}
Plasmonic nanostrctures can support localized surface plasmons (LSPs) and propagating surface plasmon polaritons (SPPs), both of which have strongly enhanced electromagnetic fields confined to deep subwavelength regions around the nanostructures. These unique properties allow for the probing of minute changes in the surrounding environment and make plamsonic nanostructures ideal platforms for realization of novel sensors for chemical and biological molecules with flexibility and enhanced performance \cite{NPhoton2014_PlasSensRev,ProcIEEE2016_PlasSensRev,CSR2017_PlasBioSensRev,NPhoton2017_PlasBioSensRev,CR2018_PlasBioSensRev}. The sensing performance can be quantified by the bulk sensitivity $S$, which is usually defined as the peak shift of the plasmonic resonance per refractive index unit, 
\begin{equation}
\label{eq:S}
S\equiv \Delta \lambda_0 / \Delta n_0\,,
\end{equation}
and the figure of merit (FOM), which is usually obtained by
dividing the bulk sensitivity by the full-width-half-maximum (FWHM), {\sl i.e.}, the linewidth of the resonance,
\begin{equation}
\label{eq:FOM}
{\rm FOM} = S/ {\rm FWHM}\,.
\end{equation}
Therefore, approaches for optimizing the plasmonic sensor' structure design include increasing the sensitivity by boosting the local field enhancement, and reducing the resonance linewidth \cite{NPhoton2014_PlasSensRev}.

Since LSP- or SPP-based sensors have relatively low sensitivities (not exceeding 300~nm/RIU) and relatively large FWHM ($\sim 60$~nm) \cite{NanoPlasSensor}, various complex plasmonic designs other than continuous metal films and isolated metal nanoparticles have been proposed or demonstrated over the years in order to enhance the sensing performance. One approach is to place two or more metal nanoparticles in close proximity with each other. For example, making use of the Fano resonance originated from the coupling between a bright mode and a dark mode, narrow spectral linewidth and enhanced local fields can be obtained, resulting in an high sensitivity up to 1000~nm/RIU and an FOM of 16/RIU \cite{ProcIEEE2016_PlasSensRev}. Another popular approach is to arrange plasmonic nanostructures in a periodic array. Kabashin {\sl et al.} \cite{NM2009_HyperBolSens} demonstrated that plasmonic nanopillar metamaterials can provide an enhanced sensitivity of 32000 nm/RIU and an FOM of 330/RIU. Yanik {\sl et al.} \cite{PNAS2011Altug_EOTsensing} made use of extraordinary optical transmission and demonstrated spectrally narrow plasmonic resonances of linewidth 4.3~nm and record high experimental FOM up to 162/RIU. Meng {\sl et al.} \cite{OL2014_SPPSens} optimized a metallic grating with shallow grooves and achieved a large sensitivity of 1400~nm/RIU, ultra-narrow FWHM of 0.4~nm at the SPP resonance wavelength of 1402~nm, and thus a huge FOM of 2300/RIU. Fang {\sl et al.} \cite{AOM2015_PlasSensHighS} demonstrated that a hybrid plasmonic metamaterial can have an ultra-high sensitivity of 4800 nm/RIU in the visible regime. Yang {\sl et al.} \cite{OE2019_SLRMIM} proposed a novel type of surface lattice resonance (SLR) with narrow FWHM that prefers the asymmetric dielectric environment, and enabled SLR-based sensing with ${\rm FOM}=28$/RIU.  

Recently, Rayleigh anomaly (RA), which is a kind of resonance only determined by the period of the grooves and the environmental refractive index, has also been used to realize plasmonic sensors. McMahon {\sl et al.} \cite{OE2007_RASens} proposed RI sensing using RA-based nanogratings. Cui {\sl et al.} \cite{ACP2012_RASens} proposed an RA-based plasmonic sensor and showed that high sensitivity of 550~nm/RIU and FOM of 1100/RIU can be numerically achieved. Savoia {\sl et al.} \cite{OE2013_RASens} investigated the surface sensitivity of RAs in metallic nanogratings. Based on the RA in plasmonic gold mushroom arrays, Shen {\sl et al.} \cite{NC2013_RAMIMSens} demonstrated a high FOM reaching $\sim108$/RIU, which is comparable to the theoretically predicted upper limit for standard propagating SPP sensors. Chen {\sl et al.} \cite{OL2018_DualRAMIMSens} made use of the RA in a metal-insulator-metal configuration and achieved a sensitivity of 1470~nm/RIU and an FOM of 6400/RIU. Su {\sl et al.} \cite{OE2019_RASens} proposed a novel grating composed of an 8-layer Au/Al$_2$O$_3$ stack and obtained narrower RA compared with the gold grating. By optimizing a metallic grating, Rahimi and Askari \cite{AO2020_RASens} obtained narrow FWHM of 0.17~nm and an ultrahigh FOM of 8530/RIU.

Quite recently, Ao {\sl et al.} \cite{ACSP2019_MirrorDSLR} demonstrated that a band edge mode near RA and an SPP can be supported in a dielectric nanopillar metasurface on an optically thick metal film. They showed that the band edge mode near RA has a linewidth of 0.5~nm and a quality factor of $Q\sim 1500$ at the resonance wavelength of 755~nm, and has greatly enhanced electric fields extending beyond the dielectric nanopillars, and that the SPP mode has a linewidth of 15~nm at the longer-wavelength of 950~nm with field localized at the bottom of the nanopillar. These exciting characteristics of the band edge mode near RA make the configuration promising for realizing refractive index sensors with high sensitivity and high FOM. Making use of this configuration, Callewaert {\sl et al.} \cite{JO2016_DonM} designed narrow band absorber based on the first order grating mode and the SPP mode, of which the linewidths are 10~nm and 27~nm, respectively; Li {\sl et al.} \cite{OE2021_SPPSens} designed an SPP-based metasurface perfect absorber with a linewidth of 15~nm at the resonance wavelength of 1348~nm, and showed that the sensitivity reaches 1067~nm/RIU. However, these linewidths is much larger than that of the band edge mode near RA in ref.~\cite{ACSP2019_MirrorDSLR}, and the RA in such a configuration has not yet been explored.

In this work, we numerically investigate the sensing performance of the high-quality RA in the mirror-backed dielectric metasurface. Strikingly, we will show that an extremely narrow linewidth of 0.045~nm can be achieved for the RA and an ultrahigh FOM of 15900 can be achieved for the RA-based sensing. These values outperform all the reported sensors based on plasmonic nanostructures. In order to understand the underlying physics, we will make use of the near-field optical picture and will discuss the effects of the dielectric nanopillar's material and size. We will also compare the sensing performance with other RA-based plasmonic sensors, and summarize the shared characteristics of RA-bases plasmonic sensors, which, we believe, will guide the structure design for better sensing performance.

\section{Theory and simulation setup}
\begin{figure}[htb]
\centering
\includegraphics[width=75mm]{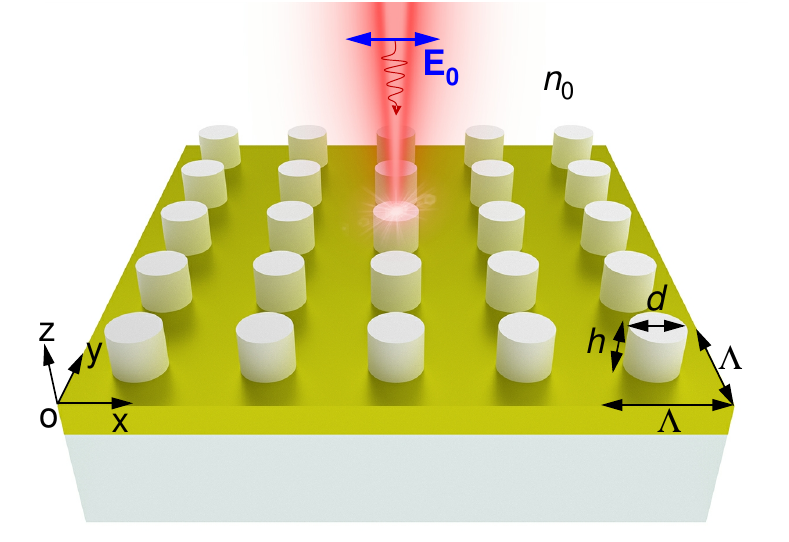}
\caption{Schematic of the metasurface under study, which is composed of a 2D ZnO nanopillar array backed by a thick gold mirror. The nanopillars have diameter $d$, height $h$, and periods $\Lambda$ in both $x$ and $y$ directions. Plane wave with electric field polarized along the $x$-axis normally impinges onto the metasurface.}
\label{fig:schem}
\end{figure}

Figure \ref{fig:schem} illustrates the mirror-backed dielectric metasurface under study, which consists of two-dimensional (2D) array of ZnO nanopillars on top of an optically thick gold film. The periodic nanopillars with diameter $d$, height $h$ and period $\Lambda$ are covered by a superstrate of refractive index $n_0$. The structure is normally illuminated by plane wave with electric field of unitary amplitude ($|E_0|=1$) and $x$-polarization. 

The metasurface can be fabricated using the state-of-the-art nanofabrication processes. It starts with the deposition of the thick gold film on a quartz substrate, followed by the deposition of a ZnO film. A photoresist is spin-coated on top and is patterned with the electron beam lithography. The pattern is then transferred to the ZnO film by etching. Finally, the remaining resist on top of the structure is removed, resulting in the 2D array of ZnO nanopillars on the gold film.

The reflectance spectra, as well as the near-field distributions of the mirror-backed dielectric metasurface were simulated with a home-developed package for fully vectorial rigorous coupled-wave analysis (RCWA), which was developed following refs.~\cite{JOSAA1995RCWA,JOSAA1997RCWA,PRB2006RCWA}. In all the simulations, we adopted retained orders of $41\times41$, which were shown to be large enough to reach the convergence regime. Wavelength-dependent optical constants of gold were taken from Johnson and Christy \cite{JC1972NK}. The influences of temperature on the material permittivities are neglected since we have the metasurface operating at room temperature. Calculations were performed with $n_0=1$, $\Lambda=720$~nm, $d=180$~nm and $h=460$~nm unless otherwise specified.

\section{Results and discussion}
\subsection{Spectrum and near fields}
\begin{figure}[htb]
\centering
\includegraphics[width=85mm]{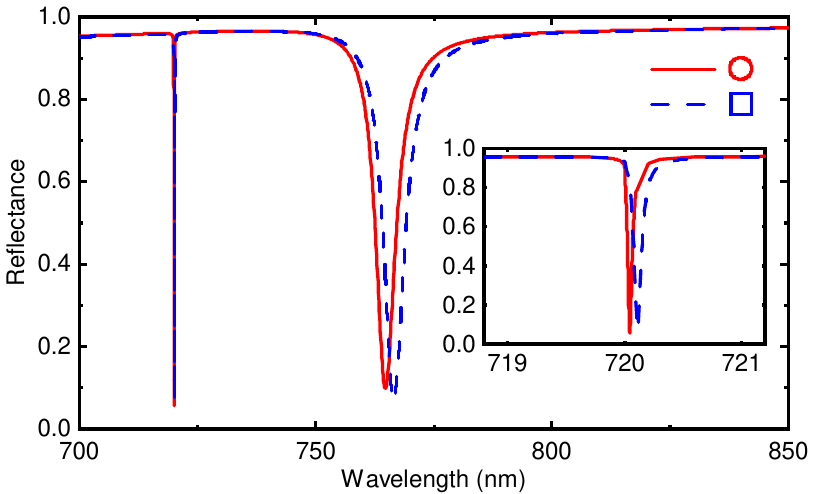}
\caption{Simulated reflectance spectra of mirror-backed dielectric metasurfaces composed of circular (red curve) or square (blue curve) nanopillars. Inset: zoom-in of the reflectance dips for the RA.}
\label{fig:spectra}
\end{figure}

Figure \ref{fig:spectra} depicts the simulated reflectance spectrum of the metasurfaces composed of circular or square nanopillars. Note that throughout this work we only present the total reflectance since it equals to the zero-order reflectance for wavelengths above the RA wavelength.  Results show find that there exists an extremely narrow Fano-shaped dip locating at $\lambda_{0} = 720.1$~nm, which is around the $(\pm 1,0)$ order RA wavelength determined by
\begin{equation}
\label{eq:RA}
\lambda_{{\rm RA}, (i,j)}= n_0 \Lambda/(\sqrt{i^{2}+j^{2}})\,.
\end{equation}
Here $(i, j)$ is the order of the RA. Remarkably, for the circular nanopillar metasurface (the red curve) the FWHM is down to 0.045~nm, and thus the quality factor calculated with $Q=\lambda_0/{\rm FWHM}$ is extremely high, reaching up to 16000. This value is an order of magnitude larger than the simulated result of the narrow band edge mode near RA in ref.~\cite{ACSP2019_MirrorDSLR}, and is about twice of the highest quality factor of the reported RA in the near infrared regime \cite{AO2020_RASens}. In contrast, for the square nanopillar metasurface (the blue curve) the FWHM of the RA is 0.08~nm, twice of that of circular structure. Here the side length of the square nanopillar is taken to be 160~nm so that the square nanopillar has similar area as the circular one. These results suggest that the roundness of the nanopillar is vital for achieving narrow RA linewidth. Hereafter, we restrict ourselves to the metasurface composed of circular nanopillars only.

Besides the extremely narrow RA dip, a relatively wider reflectance dip of Lorentzian profile due to the excitation of the SPP mode locates at $\lambda_{\rm SPP} = 764.8$~nm. The FWHM is 5.1~nm, and thus the corresponding quality factor is calculated to be $Q = 150$, which is more than twice of those of the SPP modes in \cite{ACSP2019_MirrorDSLR,OE2021_SPPSens}.

\begin{figure}[htp]
\centering
\includegraphics[width=80mm]{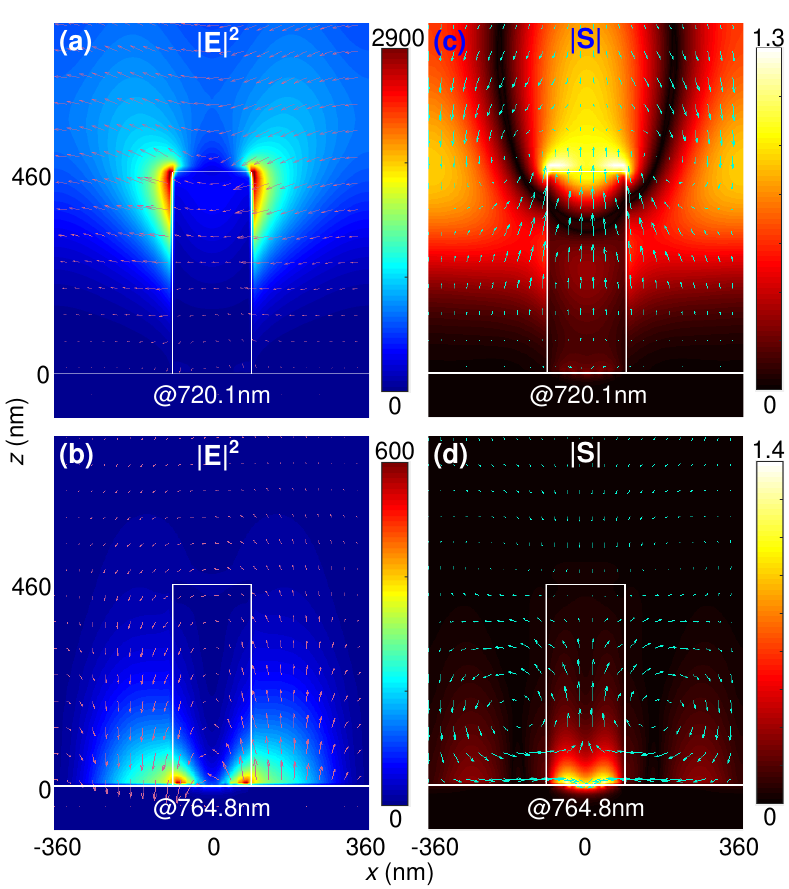}
\caption{(a)(b) Electric field intensity $|E|^2$ and (c)(d) Poynting vector $|S|$ maps (color for the intensity and arrows for the direction) in the $x-z$ plane for (a)(c) the RA at $\lambda=720.1$~nm, and (b)(d) the SPP resonance at $\lambda= 764.8$~nm.}
\label{fig:field}
\end{figure}

To understand the physics origins of the two reflectance dips, we plot the near-field electric field and Poynting vector maps in Fig.~\ref{fig:field}. Results show that at $\lambda=720.1$~nm, the electric field with well-defined direction is mainly confined to the top outside corners of the ZnO nanopillar, and the maximum electric field enhancement reaches as high as $|E|^2/|E_0|^2 = 2900$. However, at $\lambda_2 = 764.8$~nm, the electric field with vertical oscillating directions is mainly confined to the bottom inside corners of the ZnO nanopillar, and the maximum electric field enhancement is only 600. These distinct near-field distributions feature the RA and the SPP, respectively. This is consistent with the literature \cite{OE2013_RASens,ACSP2019_MirrorDSLR,OE2021_SPPSens}: for the RA, the field enhancement extends far from the metal surface \cite{OE2013_RASens}; whereas for the SPP, the electric field distribution is exponentially bound to the metal surface \cite{ACSP2019_MirrorDSLR,OE2021_SPPSens}. Correspondingly, the circulating energy flow is also far away from the metal surface for the RA, but close to the metal surface for the SPP, as shown by Figs.~\ref{fig:field}(c) and (d), respectively.

It is worth mentioning that for the RA in the mirror-backed dielectric metasurface, most electric energy with greatly enhanced fields are confined in the superstrate region, {\sl i.e.}, the sensing region, making this metasurface an ideal platform for achieving high performance sensing, as we will elaborate later. Although the electric field distribution of the RA resonance in Fig.~\ref{fig:field}(a) shares some similarities with that of the band edge mode near RA in ref.~\cite{ACSP2019_MirrorDSLR}, a major difference lies in the electric field inside the nanopillar: it is weak for the RA here but relatively strong for the band edge mode near RA in ref.~\cite{ACSP2019_MirrorDSLR}. We also note that the proposed metasurface is polarization independent due to the symmetry of the structure.

\subsection{Effects of the dielectric material and size}
Compared with \cite{ACSP2019_MirrorDSLR}, where low-index dielectric was used, in this work we adopt dielectric of larger refractive index. Here we show that the choice of the dielectric material and the size are vital for achieving the super-narrow RA.

\begin{figure}[htp]
\centering
\includegraphics[width=85mm]{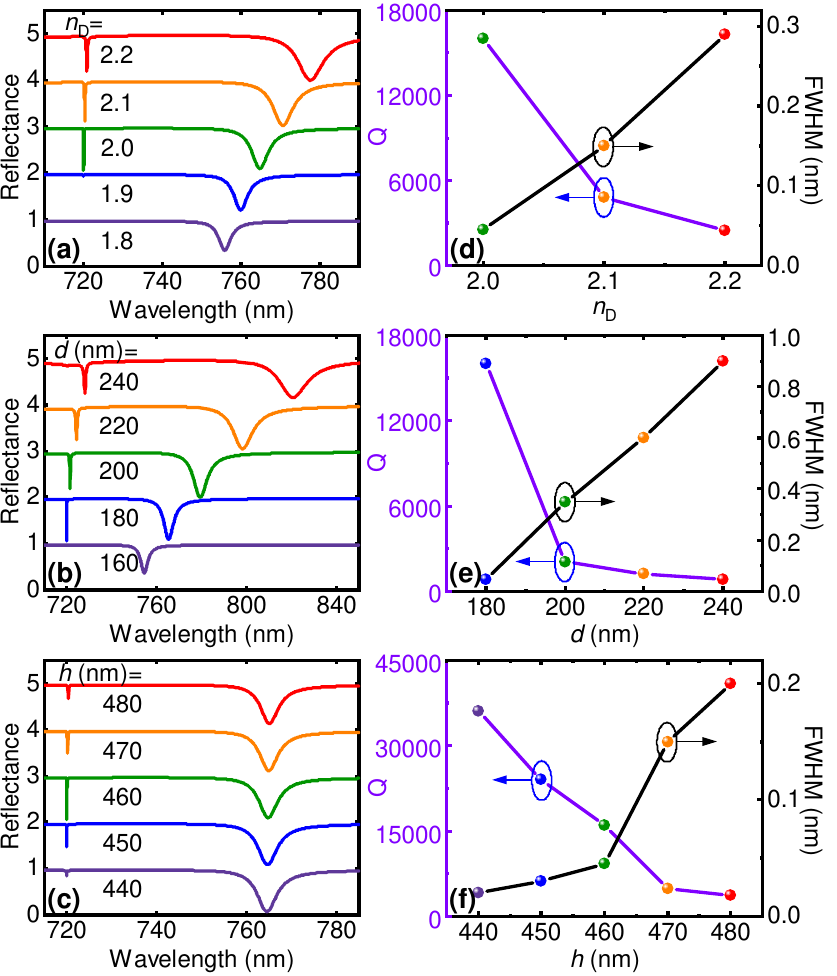}
\caption{(a)--(c) Reflectance spectra, (d)--(f) quality factors and FWHMs of RAs for different (a)(d) $n_{\rm D}$, (b)(e) $d$ and (c)(f) $h$ while all the other parameters are fixed.}
\label{fig:RvsnD}
\end{figure}

Figure \ref{fig:RvsnD}(a) depicts the evolution of the reflectance spectra as the refractive index of the dielectric nanopillar $n_{\rm D}$ varies from 1.8 to 2.2. We find that for too small refractive indices of $n_{\rm D}=1.8$ and 1.9, the RA can be hardly excited because of the low scattering efficiency of the low-index  dielectric nanopillar \cite{LSA2014scattering}. For relatively large refractive indices of $n_{\rm D}=2.1$ and 2.2, the RA reflectance dips are getting wider. Correspondingly, the FWHM increases and the quality factor decreases as $n_{\rm D}$ increases from 2.0 to 2.2, as shown by Fig.~\ref{fig:RvsnD}(d). This indicates increased loss originated from the high-dielectric/metal interface. 

Figures \ref{fig:RvsnD}(b)(c) show that, for too small nanopillar diameters or heights, the RA can be hardly excited due to the low scattering efficiency of the small dielectric nanopillar \cite{LSA2014scattering}. For larger diameters or heights, however, the FWHM increases and the quality factor decreases, as shown by Figs.~\ref{fig:RvsnD}(e)(f). These behaviors are similar to those as functions of $n_{\rm D}$ and can be explained similarly.

In Figs. \ref{fig:RvsnD}(a)(b), we also find the SPP wavelengths are red-shifted with larger FWHM as $n_{\rm D}$ or $d$ increases. This is because the fraction of energy confined within the high-index nanopillar increases. The SPP wavelength is almost independent from the nanopillar height, which is already large enough, and the FHWM slightly decreases as $h$ increases, as shown by Fig.~\ref{fig:RvsnD}(c). 

\subsection{Sensing performance}
We now study the sensing performance of the RA and the SPP in the proposed mirror-backed dielectric metasurfaces.

\begin{figure*}[hbtp]
\centering
\includegraphics[width=\linewidth]{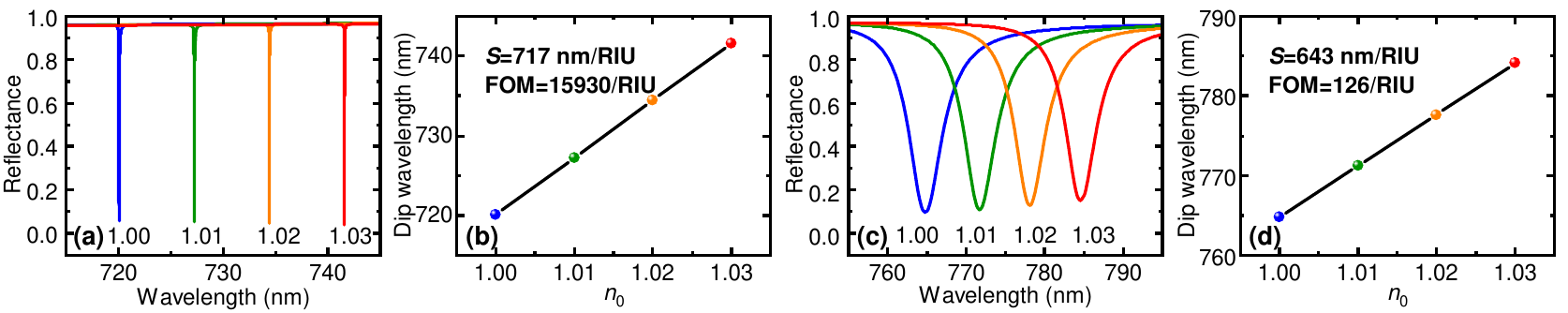}
\caption{(a)(c) Refractive-index-dependent reflectance spectra of (a) RAs and (c) SPPs. (b) RA wavelengths and (d) SPP resonance wavelengths versus the environmental refractive index.}
\label{fig:Dindex1}
\end{figure*}

Figure \ref{fig:Dindex1} shows that both the RA and the SPP resonance wavelengths are redshifted as the environmental refractive index increases. The corresponding wavelength shifts are linearly proportional to the index change. The sensitivities are calculated  to be $S=717$~nm/RIU and 643~nm/RIU for the RA- and SPP-based sensors, respectively. Their corresponding FOMs reach 15930/RIU and 126/RIU, respectively. 

\begin{figure*}[hbtp]
\centering
\includegraphics[width=\linewidth]{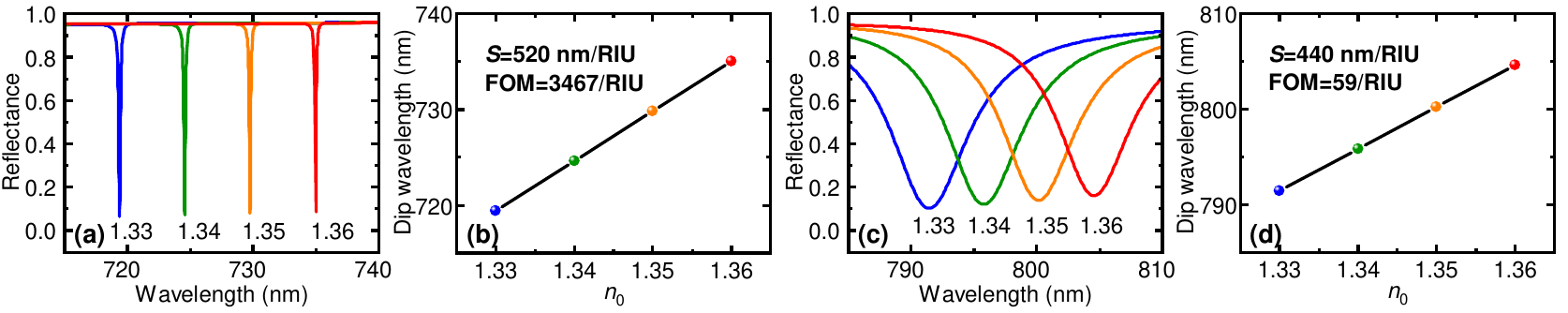}
\caption{Similar to Fig.~\ref{fig:Dindex1} but for the environmental refractive index of 1.33. The calculations were performed with optimized parameters of $\Lambda=540$~nm, $d=200$~nm, and $h=350$~nm.}
\label{fig:Dindex1p33}
\end{figure*}

Given the same RA reflectance dip wavelength, we can also design a high-$Q$ mirror-backed dielectric metasurface with environmental index of $n_0=1.33$, so that the metasurface can be used as a plasmonic sensor for biological molecules. With $n_0=1.33$ and $\lambda_0\approx 720$~nm, the optimized size parameters are $\Lambda=540$~nm, $d=200$~nm, and $h=350$~nm. In this scenario, Fig.~\ref{fig:Dindex1p33}(a) shows that the RA locating at $\lambda_0=719.4$~nm also has a narrow FWHM of 0.15~nm, corresponding to a quality factor of $Q=4796$. As the environmental index increases, Fig.~\ref{fig:Dindex1p33} also shows red shifts of the RA and the SPP resonance wavelengths. The sensitivities are 520~nm and 440~nm, and the corresponding FOMs are 3467/RIU and 59/RIU, respectively.

\begin{table*}[htb]
 \caption{\label{tab:Comp}Comparison on performance of RA-based plasmonic sensors. The units of $\lambda_0$, $S$, $S^*$, FWHM, and FOM are nm, nm/RIU, 1/RIU, nm, and 1/RIU, respectively.}
 \centering
 \begin{tabular}{ccccccccc}
 \hline
 $\lambda_0$ &$n_0$ &$S$ &$S^*$ &FWHM &$Q$ &FOM   &$n_0 S^*$   & Reference\\
   \hline
1288 &1.33 &1015 &0.788 &9.5 &136 &$\sim 108$  &1.048   &\cite{NC2013_RAMIMSens} \\
1477.1 &1.0 &1470 &0.995 &0.23 &6422 &6400  &0.995   &\cite{OL2018_DualRAMIMSens} \\
666 &1.33 &496 &0.745 &1.8 &370 &275  &0.991
&\cite{OE2019_RASens}\\
1530 &1.0 &1450 &0.948  &0.17 &9000 &8530  &0.948
&\cite{AO2020_RASens} \\
720.1 &1.0 &717 &0.996  &0.045 & {\bf 16000} &{\bf 15930 }  &0.996 &{\bf This work} \\
719.4 &1.33 &520 &0.723 &0.15 &{\bf 4796} &{\bf 3467}   &0.963  &{\bf This work} \\
\hline
\end{tabular}
\end{table*}

Table~\ref{tab:Comp} summarizes a short survey of the literature on RA-based plasmonic sensors. It is clear that for both scenarios of $n_0=1$ and $n_0=1.33$, our metasurfaces provide the highest quality factors and the highest FOMs, which are one order of magnitude larger than the largest values in the literature. Moreover, because the ohmic loss can be further reduced when the RA wavelength is tuned from the visible to the near-infrared region, which can be done by using larger lattice periods according to Equation~(\ref{eq:RA}), we expect further increases of the quality factor and the FOM.    

In Table~\ref{tab:Comp} we find that the sensitivities of all the RA-based plasmonic sensors are wavelength-dependent, and to be more specifically, are proportional to the resonance wavelength, consistent with other plasmonic sensors \cite{Langmuir2013_PlasSensRev,NanoPlasSensor}. From Equation~(\ref{eq:S}), we have
\begin{equation}
\label{eq:S2}
S = \frac{\partial \lambda_0}{\partial n_0} \approx \frac{\partial \lambda_{{\rm RA},(i,j)}}{\partial n_0} = \frac{\lambda_{{\rm RA},(i,j)}}{n_0} \approx \frac{\lambda_{0}}{n_0}\,,
\end{equation}
where $\lambda_0 \approx \lambda_{{\rm RA},(i,j)}$ has been included. We also introduce the concept of {\sl normalized sensitivity}, which is defined as
\begin{equation}
\label{eq:Snorm}
S^* \equiv S/\lambda_{0}\,.
\end{equation}
Combining Equations~(\ref{eq:S2}) and (\ref{eq:Snorm}), we have
\begin{equation}
\label{eq:nSnorm}
n_0 S^* \approx 1\,.
\end{equation}
From Table~\ref{tab:Comp} we find that $S^*\approx 1$ for $n_0=1$ and $S^*\approx 0.75$ for $n_0=1.33$, both satisfying Equation~(\ref{eq:nSnorm}).
Therefore, for the RA-based plasmonic sensors, both the theory and the numerical results have revealed that the sensitivity $S$ and the normalized sensitivity $S^*$ are well defined by $n_0$ and $\lambda_0$. In other words, given the environmental index and the operation wavelength, which are determined by the sensing application (for example, $n_0=1$ for gas sensing and $n_0=1.33$ for biomedical sensing) and the laser used, the sensitivity or the normalized sensitivity of RA-based sensors can be hardly be improved.

Combining Equations~(\ref{eq:FOM}) and (\ref{eq:S2}), we obtain
\begin{equation}
\label{eq:FOM2}
{\rm FOM} \approx Q/n_0\,.
\end{equation}
This means that the FOM of the RA-based plasmonic sensor is defined only by the quality factor and the environmental index, as verified by Table~\ref{tab:Comp}. Therefore, according to Equations~(\ref{eq:S2}) and (\ref{eq:FOM2}), for the RA-based plasmonic sensor with given $n_0$ and $\lambda_0$, the sensitivity cannot be improved regardless of the structure design, and only the FOM or equivalently the quality factor can be optimized. In that sense, the mirror-backed dielectric metasurface outperforms all the other reported plasmonic nanostructures.

\section{Conclusions}
In conclusion, we have numerically shown that the mirror-backed dielectric metasurface consisting of periodic ZnO nanopillars on top of a gold film can support RAs with an ultrahigh quality factor reaching 16000, and that the RA-based sensor can have an extremely high FOM of 15930. These values are an order of magnitude larger than the highest values in the literature. We have attributed these striking performances to the reduced ohmic loss since the strongly enhanced electric fields are far away from the metal film. By investigating the effects of the dielectric material and the size, we have shown that there exists a subtle balance between the scattering efficiency and the absorption loss, suggesting that these parameters should be carefully optimized for achieving super-high quality factors. By summarizing the performances of RA-based plasmonic sensors, we have revealed their shared characteristics: the sensitivity is approximate to the ratio of the operation wavelength and the environmental index, and the FOM is approximate to the ratio of the quality factor and the environmental index. Therefore, the sensitivity of an RA-based plasmonic sensor with given environmental index and operation wavelength is well defined, and only the FOM can be improved through the structure design. We expect these findings will advance the development of RA-based plasmonic sensors, and the proposed high-performance mirror-backed dielectric metasurface sensor will find applications in biochemical sensing.


\section*{Acknowledgments}
This work was supported by the State Key Laboratory of Advanced Optical Communication Systems and Networks, China (2020GZKF004, 2019GZKF2).

\bibliographystyle{unsrt}
\bibliography{sample}

\end{document}